\PassOptionsToPackage{unicode}{hyperref}
\PassOptionsToPackage{hyphens}{url}
\documentclass[
]{article}
\usepackage{xcolor}
\usepackage[margin=1in]{geometry}
\usepackage{amsmath,amssymb}
\usepackage{iftex}
\ifPDFTeX
  \usepackage[T1]{fontenc}
  \usepackage[utf8]{inputenc}
  \usepackage{textcomp} % provide euro and other symbols
\else % if luatex or xetex
  \usepackage{unicode-math} % this also loads fontspec
  \defaultfontfeatures{Scale=MatchLowercase}
  \defaultfontfeatures[\rmfamily]{Ligatures=TeX,Scale=1}
\fi
\usepackage{lmodern}
\ifPDFTeX\else
\fi
\IfFileExists{upquote.sty}{\usepackage{upquote}}{}
\IfFileExists{microtype.sty}{% use microtype if available
  \usepackage[]{microtype}
  \UseMicrotypeSet[protrusion]{basicmath} % disable protrusion for tt fonts
}{}
\makeatletter
\@ifundefined{KOMAClassName}{% if non-KOMA class
  \IfFileExists{parskip.sty}{%
    \usepackage{parskip}
  }{% else
    \setlength{\parindent}{0pt}
    \setlength{\parskip}{6pt plus 2pt minus 1pt}}
}{% if KOMA class
  \KOMAoptions{parskip=half}}
\makeatother
\usepackage{color}
\usepackage{fancyvrb}

\DefineVerbatimEnvironment{Highlighting}{Verbatim}{commandchars=\\\{\}}
\usepackage{framed}
\definecolor{shadecolor}{RGB}{248,248,248}
\newenvironment{Shaded}{\begin{snugshade}}{\end{snugshade}}

\newcommand{\CommentTok}[1]{\textcolor[rgb]{0.56,0.35,0.01}{\textit{#1}}}

\newcommand{\FunctionTok}[1]{\textcolor[rgb]{0.13,0.29,0.53}{\textbf{#1}}}

\newcommand{\NormalTok}[1]{#1}

\newcommand{\OtherTok}[1]{\textcolor[rgb]{0.56,0.35,0.01}{#1}}

\usepackage{graphicx}
\makeatletter
\newsavebox\pandoc@box
\newcommand*\pandocbounded[1]{% scales image to fit in text height/width
  \sbox\pandoc@box{#1}%
  \Gscale@div\@tempa{\textheight}{\dimexpr\ht\pandoc@box+\dp\pandoc@box\relax}%
  \Gscale@div\@tempb{\linewidth}{\wd\pandoc@box}%
  \ifdim\@tempb\p@<\@tempa\p@\let\@tempa\@tempb\fi% select the smaller of both
  \ifdim\@tempa\p@<\p@\scalebox{\@tempa}{\usebox\pandoc@box}%
  \else\usebox{\pandoc@box}%
  \fi%
}
\def\fps@figure{htbp}
\makeatother
\NewDocumentCommand\citeproctext{}{}

\makeatletter
 \let\@cite@ofmt\@firstofone
 \def\@biblabel#1{}
 \def\@cite#1#2{{#1\if@tempswa , #2\fi}}
\makeatother
\newlength{\cslhangindent}
\newlength{\csllabelwidth}
\newenvironment{CSLReferences}[2] % #1 hanging-indent, #2 entry-spacing
 {\begin{list}{}{%
  \setlength{\itemindent}{0pt}
  \setlength{\leftmargin}{0pt}
  \setlength{\parsep}{0pt}
  \ifodd #1
   \setlength{\leftmargin}{\cslhangindent}
   \setlength{\itemindent}{-1\cslhangindent}
  \fi
  \setlength{\itemsep}{#2\baselineskip}}}
 {\end{list}}
\usepackage{calc}

\providecommand{\tightlist}{%
  \setlength{\itemsep}{0pt}\setlength{\parskip}{0pt}}
\usepackage{bookmark}
\IfFileExists{xurl.sty}{\usepackage{xurl}}{} % add URL line breaks if available
\hypersetup{
  pdftitle={HybridQC: Machine Learning-Augmented Quality Control for Single-Cell RNA-seq Data},
  pdfauthor={Kaitao Lai1},
  hidelinks,
  pdfcreator={LaTeX via pandoc}}

\title{HybridQC: Machine Learning-Augmented Quality Control for
Single-Cell RNA-seq Data}
\author{Kaitao Lai\textsuperscript{1}}
\date{2025-09-03}

\begin{document}
\maketitle

\textsuperscript{1} University of Sydney

\section{Summary}\label{summary}

HybridQC is an R package that streamlines quality control (QC) of
single-cell RNA sequencing (scRNA-seq) data by combining traditional
threshold-based filtering with machine learning-based outlier detection.
It provides an efficient and adaptive framework to identify low-quality
cells in noisy or shallow-depth datasets using techniques such as
Isolation Forest (Liu et al. 2018), while remaining compatible with
widely adopted formats such as Seurat objects.

The package is lightweight, easy to install, and suitable for
small-to-medium scRNA-seq datasets in research settings. HybridQC is
especially useful for projects involving non-model organisms, rare
samples, or pilot studies, where automated and flexible QC is critical
for reproducibility and downstream analysis.

\section{Statement of Need}\label{statement-of-need}

scRNA-seq experiments often suffer from technical noise, dropout events,
and variability in sequencing depth. Traditional quality control relies
on static cutoffs for metrics such as gene count, UMI count, and
mitochondrial content, which may be suboptimal for non-standard datasets
(Malte D. Luecken and Theis 2022).

HybridQC fills this gap by integrating machine learning
methods---specifically unsupervised outlier detection with Isolation
Forest (Liu et al. 2018)---to improve filtering precision and
robustness. This dual-level approach can better preserve informative but
unconventional cell types and adapt dynamically to diverse datasets. No
existing R packages provide this hybrid QC strategy as a standalone tool
with seamless integration into Seurat-based pipelines.

\section{Features}\label{features}

\begin{itemize}
\tightlist
\item
  Computes standard QC metrics: \texttt{nFeature\_RNA},
  \texttt{nCount\_RNA}, \texttt{percent.mt}
\item
  Supports Isolation Forest outlier detection via \texttt{reticulate}
  and \texttt{pyod} (Liu et al. 2018)
\item
  Filters cells using a hybrid decision rule
\item
  Works on Seurat objects
\item
  Lightweight and suitable for quick prototyping or small studies
\end{itemize}

\section{Example Usage}\label{example-usage}

We demonstrate HybridQC on a synthetic single-cell RNA-seq dataset
derived from 10x Genomics-style PBMC data, consisting of 2,000 cells and
1,000 genes. The dataset is loaded into R as a Seurat object and
subjected to a two-stage quality control workflow.

First, we compute basic quality metrics such as gene count, UMI count,
and mitochondrial content per cell. Then, an unsupervised outlier
detection model (Isolation Forest) is applied to capture multivariate
anomalies not addressed by static thresholds alone. Cells flagged by
either criterion are filtered out, resulting in a cleaner, more reliable
dataset for downstream analysis.

\begin{Shaded}
\begin{Highlighting}[]
\FunctionTok{library}\NormalTok{(Seurat)}
\FunctionTok{library}\NormalTok{(HybridQC)}

\NormalTok{pbmc }\OtherTok{\textless{}{-}} \FunctionTok{LoadPBMC2k}\NormalTok{()  }\CommentTok{\# Load example data}
\NormalTok{qc\_basic }\OtherTok{\textless{}{-}} \FunctionTok{run\_basic\_qc}\NormalTok{(pbmc)}
\NormalTok{ml\_scores }\OtherTok{\textless{}{-}} \FunctionTok{run\_isolation\_forest\_qc}\NormalTok{(pbmc)}
\NormalTok{filtered }\OtherTok{\textless{}{-}} \FunctionTok{filter\_cells}\NormalTok{(pbmc, qc\_basic, ml\_scores)}
\end{Highlighting}
\end{Shaded}

\section{UMAP Visualization of Outlier
Scores}\label{umap-visualization-of-outlier-scores}

To visualize Isolation Forest--based anomaly scores in a low-dimensional
embedding, HybridQC supports \texttt{FeaturePlot()} using UMAP
projections (McInnes, Healy, and Melville 2018). The resulting plot
highlights which cells are predicted as outliers in UMAP space based on
their multivariate QC metrics.

Below is an example UMAP colored by Isolation Forest score:

\begin{figure}
\centering
\pandocbounded{\includegraphics[keepaspectratio,alt={UMAP visualization of Isolation Forest outlier scores}]{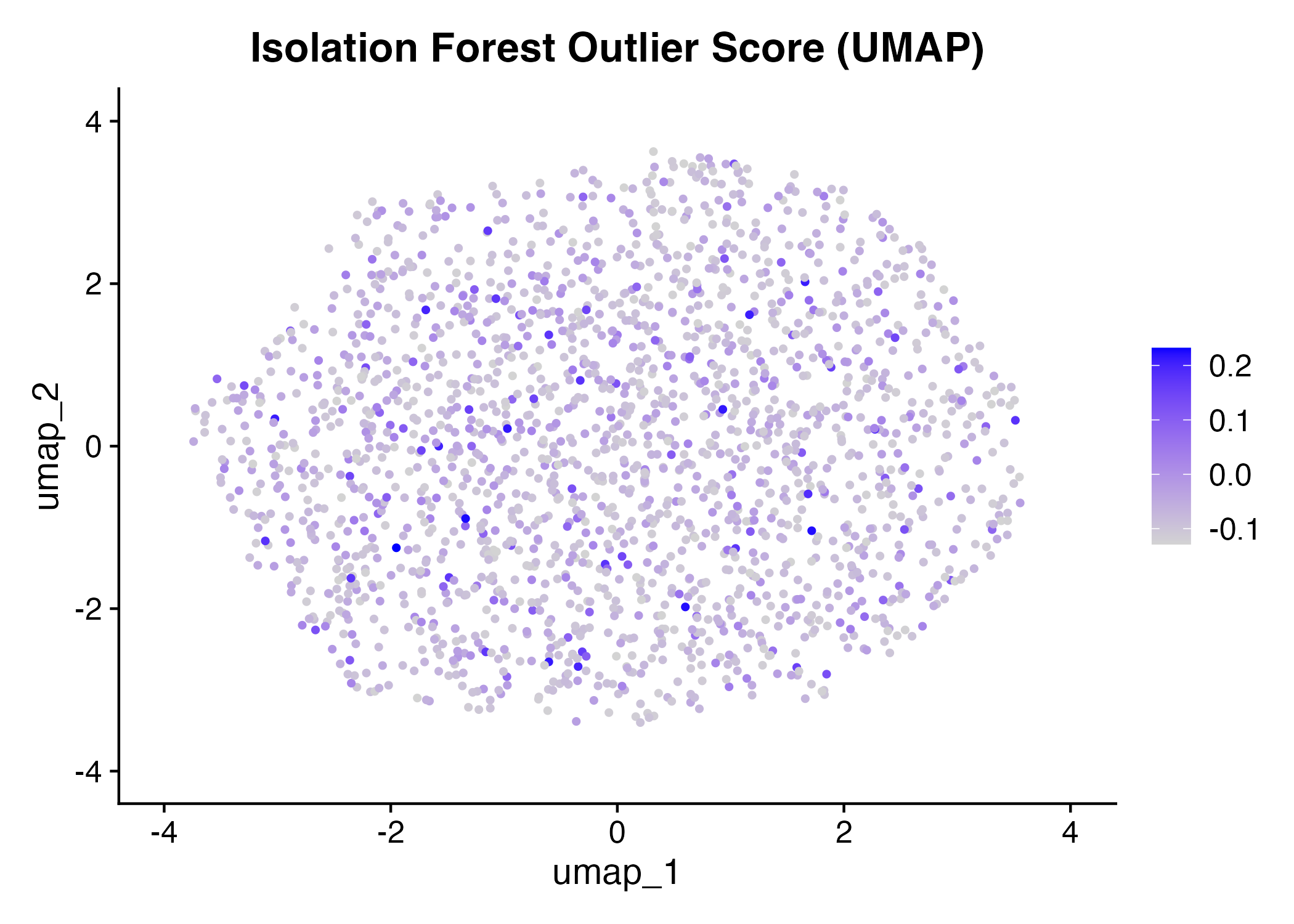}}
\caption{UMAP visualization of Isolation Forest outlier scores}
\end{figure}

\section{Software Repository}\label{software-repository}

The source code for HybridQC is freely available on GitHub at:\\
\url{https://github.com/biosciences/HybridQC}

\section{Acknowledgements}\label{acknowledgements}

The author thanks collaborators at University of Sydney for feedback on
early concepts.

\section*{References}\label{references}
\addcontentsline{toc}{section}{References}

\protect\phantomsection\label{refs}

\end{document}